\documentclass[12pt]{article} 
\usepackage{epsf}

\begin{document}
\begin{center}
{\bf \LARGE Screened Potential of a Moving Meson In a Quark Gluon Plasma}
\end{center}
\vspace*{1cm}
\begin{center}
{\bf \Large Sidi C. Benzahra \footnote[1] {benzahra@physics.umn.edu}}
\end{center}
\begin{center}
{\large School of Physics and Astronomy}
\end{center}
\begin{center}
{\large University of Minnesota}
\end{center}
\begin{center}
{\large Minneapolis, MN 55455}
\end{center}
\vspace*{1cm}

\begin{abstract}
\noindent
We consider the quark antiquark potential of a quarkonium moving with velocity $v$ through a quark-gluon plasma at 
temperature T.  An explicit, configuration-space expression is found for the screened interaction between the quarks 
constituting the meson.  This potential is non-spherical, but axially symmetric about the direction of $v$.     
\end{abstract}

Here we consider the bottom quark as a test charge Q moving in a quark-gluon plasma with velocity $v$ on the z axis. Since 
the bottom quark is heavy, we can ignore the energy transfer between the test charge and the quark-gluon plasma. This
implies that in the rest frame of the bottom quark, the field is static and can be found by calculating Poisson's equation 
for the static potential $A^{\mu}(\bf r)$[1]:
 
\begin{equation}
{\bf \Delta} A^{\mu}({\bf r}) = j^{\mu}_{ext}({\bf r}) + j^{\mu}_{ind}({\bf r}) \, ,
\end{equation} 
where $j^{\mu}_{ext}=-Q\delta({\bf r})(1,0,0,0)$ is the external current associated with the test charge placed at ${\bf
r}=0$ and $j^{\mu}_{ind}$ is the induced (static) current in the medium which is relatively moving with velocity ${\bf \rm v}
=(0,0,v)$.  In contrast with the usual Debye screening where only the scalar potential $\phi=A^{0}$ appears, one should
introduce the vector potential since the test charge will induce not only the polarization charge density but also a stable
polarization charge current in the relatively moving quark-gluon plasma.  This current generates vector potentials which
give rise to a magnetic field around the test charge.  We list below the final result, given by [1], for one of the Fourier
components of the potential discussed above:

\begin{equation}
\tilde{A}^{0}(k)=2 \pi Q \delta (\omega)\left [{{1-\gamma^{2}(1-z^{2})}\over { {\bf k}^{2} \epsilon_{T}({\bf k})}}+ 
 {{\gamma^{2}(1-z^{2})}\over { {\bf k}^{2} \epsilon_{L}({\bf k})}}\right] \, ,
\end{equation}

where 

\begin{equation}
\epsilon_{T}({\bf k})=1+{{{m^{2}_{e}}\over {{\bf k}^2}}} \phi_{T}(z) \quad ; \quad  \epsilon_{L}({\bf k})=1+{{{m^{2}_{e}}
\over {{\bf k}^2}}} \phi_{L}(z)
\end{equation}
are the transverse and longitudinal dielectric \, ``constants.'' $m_{e}$ is the inverse screening length, $\phi_{T}$ and 
$\phi_{L}$ are complex functions written in terms of $z$ only--see reference [1]--and $z$ is defined by 

\begin{equation}
z=\left [ {{k.u}\over{\sqrt{(k.u)^2 -k^2}}} \right ]_{\omega =0} = {{v {\rm cos}\theta}\over {\sqrt{1-v^2 {\rm sin}^2
\theta}}}
\end{equation} 

where $u^{\mu}=\gamma (1,0,0,v)$ and $k^{\mu}=(\omega, | {\bf k}| {\rm cos}\phi {\rm sin} \theta,  
|{\bf k}| {\rm sin}\phi  {\rm sin} \theta, | {\bf k}|{\rm cos}\theta)$, and $k.u \equiv k^{\mu} u_{\mu}.$  Knowing this, and
having the Fourier component of the potential, we can calculate the potential in a cylindrical coordinate system.  It
will take the form
\begin{equation}
A^{0}(r)={{Q m_{e}}\over {4 \pi}}F(m_{e} \rho, m_{e} z) \, .
\end{equation}

This potential, where F is a dimentionless function, is the sum of two screened potentials created by $Q$ and $\bar{Q}$.
In the Fourier transformation of eq.(2), both the real and the imaginary parts of the momentum-space potential
$\tilde{A}^{0}(k)$ contribute to the real part of $A^{0}(\rho, z)$ but no imaginary part appears in the configuration space
potential. Upon the substraction of the divergent self-energy terms, this results in

\begin{equation}
V_{e} (r) ={1 \over 2} [Q A_{ \bar{Q}^{0}} (-r) + \bar{Q} A^{0}_{Q} (r) ] \, . 
\end{equation}

This potential is difficult to calculate analytically, but since the upsilon meson is heavy and its velocity in the 
quark-gluon plasma is relatively small, the potential in configuration space can be analytically determined. At small 
velocity, the transverse and longitudinal dielectric \, ``constant \, '' of eq.(3) take the forms

\begin{equation}
\epsilon_{T}\approx 1-i {{m_{e}^{2}\pi v {\rm cos \theta} }\over {4 k^{2}}}  \quad ; \quad  \epsilon_{L}\approx (1+{{m_{e}^{2}} \over {k^{2}}}) 
+i {{m_{e}^{2}\pi v {\rm cos \theta} }\over {2 k^{2}}} \, .
\end{equation}

The potential in cylindrical coordinate will be composed of two terms.  The first term will have the Debye potential as 
it is expected, but the second term will be an extra potential that arose out of the motion of the meson. Because the meson 
is moving in the z direction, we can fix one coordinate and obtain  $\vec{k}.\vec{r}=k {\rm cos \phi sin \theta \rho} + kz 
\rm{cos \theta}.$ Using the Fourier transform to evaluate the two-body potential between a pair of test charges, and 
considering the velocity to be small, we find

\begin{eqnarray}
V & = & 2{{Q\bar{Q}}\over {(2\pi)^2}} \int_{0}^{\infty} dk \int_{0}^{1} {\rm
cos}(kzx)J_{0}(k \rho \sqrt{1-x^2}) \times \\ 
  &   & \left [  {{1}\over{(k^2
+m_{e}^2)}}-{{m_{e}^4 \pi^2 v^2 x^2} \over {4 (k^2 +m_{e}^2)^3}} \right ] k^2 dx \, ,
\nonumber
\end{eqnarray}  
where $x=\rm cos \theta$. If we calculate the last equation above, and replace the $Q \bar{Q}/(4\pi)$ by 
the color-singlet strength of the one-gluon-exchange force $4 \alpha_{s}/3$, we get a potential of the form

\begin{equation}
V=V_{0}(r)+V_{2}(r)Y_{0}^{2}(\theta, \phi) \, ,
\end{equation}

where the first part include the Debye potential

\begin{equation}
V_{0}(r)=-{{4 \alpha_{s}} \over {3}} {{e^{-m_{e} r}} \over {r}} + v^2 \alpha_{s}  {{\pi^2}\over{72}} e^{- m_{e} r} m_{e} 
(1+m_{e} r) \, .
\end{equation}

and the second part of the potential takes the form of
\begin{equation}
V_{2}(r)Y_{0}^{2}(\theta, \phi)={{{\alpha_{s} v^2 \pi^2} \over {12}}e^{-m_{e}r}Y_{0}^{2}(\theta, \phi) \sqrt{{4 \pi }\over{5}}\left [ {8\over{m_{e}r^2}}+{4\over{r}}+{8\over{m_{e}^2
r^3}}-{{8 e^{m_{e} r}}\over{m_{e}^{2} r^{3}}}+{{4 m_{e}}\over{3}}+{{m_{e}^{2} r}\over{3}} \right ]}
\end{equation} 
which can also be written in a simple form
\begin{equation}
V_{2}(r)Y_{0}^{2}(\theta, \phi)={{{\alpha_{s} v^2 \pi^2} \over {12}}e^{-m_{e}r}Y_{0}^{2}(\theta, \phi) 
\sqrt{{4 \pi }\over{5}}{8\over{m_{e}^{2}r^3}} \left [ \Sigma_{n=0}^{4} {{(m_{e}r)^{n}}\over{n!}}-e^{m_{e}r}   \right ]} \, .
\end{equation} 
In the $r\longrightarrow 0$ limit, this potential vanishes like $r^{2}$.  This is very exciting because this potential
agrees with the fact that when the quark and the antiquark are close to each other, they will not ``see'' each other due to
asymptotic freedom.  If we plotted the equipotential surfaces of $V$ 
in the $\rho$-$z$ plane, we would see that when the velocity of the meson vanished, the shape of these equipotential 
surfaces would become identical to that of the screened Coulomb potential, but if the velocity increased, the 
equipotential surfaces would become narrow in the longitudinal direction.\\

\begin{center}
{\bf \large References}
\end{center}
[1] M.C. Chu and T. Matsui, Physical Review D 39, 1892 (1989).

\end{document}